\documentclass[a4paper,11pt]{article}

\usepackage[pdftex]{graphicx}
\usepackage[utf8]{inputenc}
\usepackage[square]{natbib}

\newcommand{\deq}{\stackrel{\rm {\mathrm{def}}}{=}}

\newcommand{\nucleus}{\mathcal{N}}

\title{
Inspiration from genetics to promote recognition and protection within {\em ad hoc} sensor networks
}

\author{
Reinert Korsnes
\and 
Knut {\O}vsthus\thanks{Bergen University College,
 P.O.\ Box 7030,
 N-5020 Bergen, Norway}
}

\begin{document}
\maketitle

\begin{abstract}
This work illustrates potentials for recognition within {\em ad hoc} sensor networks if their nodes possess individual inter-related biologically inspired genetic codes.
The work takes ideas from natural immune systems protecting organisms from infection.
Nodes in the present proposal have individual gene sets fitting into a self organised phylogenetic tree.
Members of this population are genetically ''relatives''.
Outsiders cannot easily copy or introduce a new node in the network without going through a process of conception between two nodes in the population. 
Related nodes can locally decide to check each other for their genetic relation without directly revealing their gene sets.
A copy/clone of a gene sequence or a random gene set will appear as alien. 
Nodes go through a cycle of introduction (conception or ''birth'') with parents in the network and later exit from it (''death''). 
Hence the phylogenetic tree is dynamic or possesses a genetic drift.
Typical lifetimes of gene sets and number of offspring from different parents affect this genetic drift and the level of correlation between gene sets.
The frequency of mutations similarly affects the gene pool.
Correlation between genes of the nodes implies a common secret for cryptographic material for communication and consistency check facilitating intrusion detection and tracing of events.
A node can by itself (non-specifically) recognise an adversary if it does not respond properly according to its genes.
Nodes can also collaborate to recognise adversaries by communicating response
from intruders to check for consistency with the whole gene pool (phylogenetic tree).
\end{abstract}

\section{Self-protecting networks and robustness}
This work takes inspiration from natural immune systems to obtain self-organised recognition and protection within {\em ad hoc} sensor networks \cite[]{Akyildiz2002}. 
The assumed threat image here is introduction of {\em false} (hostile) nodes in a network to monitor traffic and to corrupt the systems. 
Immune systems have for vertebrates both an innate (non-specific and static) and an adaptive part which recognise special characteristics of pathogens.
This protection system include recognition of genetically self and non-self.
\cite{Aickelin2007,Jungwon2007,Akyildiz2002} provide a review of concepts for artificial immune systems to protect computer systems.
A reason for this development is the apparent weakness within traditional computer security systems.
Security concerns increasingly affect management of computerised systems.
A survey by FBI/CSI shows that 34 percent of the respondent organisations spent more than 5 percent of their total IT budget on IT-security in 2006 \cite{Gordon2006,Shafi2007}.
One can therefore expect a variety of approaches within this field.

Robustness is one of the fundamental characteristics of biological systems \cite[]{Kitano2007}.
Kitano \cite{Kitano2004} defines robustness as a property that allows a system to maintain its functions against internal and external perturbations.
Stelling {\em et al.} \cite{Stelling2004} similarly phrase that ''{\em robustness, the ability to maintain performance in the face of perturbations and uncertainty, is a long-recognised key property of living systems}''.
Functionality within computer systems often depends on security measures and their robustness. 
Observed fault tolerance of biological systems is a good reason to seek inspiration from biology when considering security for complex and autonomous network systems \cite{Tampesti2007}.
Complex computer systems typically result from a development driven by empirical work where not anticipated problems are managed on an {\em ad hoc} basis \cite[]{Somayaji2007}.
This tends to make computer systems similar to biological systems which are complex and process information self-organised and distributed.

Several authors have pointed out system similarities to biology to clarify computer security issues.
The well known concept ''computer virus'' directly refers to similarities between computers and biological systems \cite[][]{Cohen1987,Guinier1989}.
Li and Knickerbockera \cite{Li2007} point out similarities between computer worms and biological pathogens.
They found that successful computer viruses typically share common tactics as found for biological pathogens. 
Shafi and Abbass \cite{Shafi2007}and Somaya \cite{Somayaji2007} survey attempts to secure complex computer systems by biologically-inspired adaptive systems and immunology.
Ibrahim and Maarof \cite{Ibrahim2005} also review biologically inspired approaches to cryptology.

It is common experience form biology that recognition is important for protection.
Examples are insect and cell communities.
Social communities and higher order animals also exercise protective recognition and individualism.
Computer systems may similarly obtain self-defense.

\section{Illustration by simulation examples}
\label{se:example}

\subsection{Main purpose and result from present simulations}
\label{se:simulation_philosophy}
The following simulation examples illustrate generation of gene sets for nodes in an assumed {\em ad hoc} stationary sensor network. 
These examples show that the correlation between the gene sets of different nodes for the present approach can stay within a range which is sufficient for the genes to serve as distributed cryptographic material for the network.
The generation and distribution of cryptographic material here take place as a by-product of uncorrelated interaction between couples of nodes. These contacts make for example mutations to diffuse throughout and renew the gene pool.
The present examples are only meant for communicating ideas about potentials for self-organised locally based protection. 
Caution must therefore be exercised not to confuse these illustrations with quantitative analysis or design.

\subsection{System initiation}
\label{se:initiation}
Assume an {\em ad hoc} sensor network starts to increase from two nodes until a fixed number of nodes (a stationary sensor network). 
This ''bootstrap'' process can take place via a protected channel (for example before physical deployment).
Fast establishment of network may also reduce possibilities for attacks (reduce vulnerability) since there may be unlikely that a possible adversary maintains constant monitoring or readiness within the actual area.
An option to protect the network initiation is application of initial cryptographic codes.
Nodes which join the network during this initial period, form genetic codes by receiving a mix of genetic codes from parents within the network.
Random ''mutations'' enter this mix of genes.

\subsection{Development of gene pool}
\label{se:development_of_gene_pool}
The present results are form simulations of 100 and 1000 nodes in a network. 
The authors made the simulation tool by direct programming in Ada 2005 (GNU Ada under Linux). 

Each node in the simulations stores a data sequence similar to a gene in a biological organism.
An indexed set of variables $G_{i}^{X}, i = 1,\ldots,n$ represents a gene sequence of a node $X$ where $n = 1000$. 
For each $i = 1,2,\ldots,n$, the ''nucleotide'' $G_{i}^{X}$ has values in the range $\nucleus \deq \{ A, B, \ldots, Z \}$. 
Two nodes $X_{1}$ and $X_{2}$ can make an offspring $X_{3}$ via merging their genes so that $G_{i}^{X_{3}} = G_{i}^{X_{1}}$ and $G_{i}^{X_{3}} = G_{i}^{X_{2}}$ with equal probability $p = 0.48$
and a probability $p_{\mathrm{m}} = 0.04$ for a random mutation $G_{i}^{X_{3}} \in \nucleus$
with uniform distribution.

A node enters the network via a random selection of two parents to make an offspring as described above.
The population of linked nodes increases from two initial nodes (''parents'') up to the maximum of 100 nodes (1000 nodes for the second simulation). 
A random node exits this population (''dies''), when the number of nodes exceeds this number.
It later returns according to the procedure above. 
Table \ref{tab:genes1} shows an example of gene sequences (the 70 first parts). 
\renewcommand{\arraystretch}{0.85}
\begin{table}[htp]
\caption{Example of gene sequences.}
\label{tab:genes1}
\begin{center}
\resizebox{\textwidth}{!}{
\texttt{
\begin{tabular}{|c|l|} \hline
 \mbox{} &  \\
 \multicolumn{1}{|l|}{{\bf
 Node}} & \multicolumn{1}{c|}{{\bf Gene sequence (with bases A..Z)}}    \\ 
 \mbox{} &  \\ 
 1   & OBERJRDALRIWDABIPIAQATTLVBLMRSMAUSKVAAWFXJVVMVFERZTTUDEYMKMCKQSXLRBUHVQDWLFZAXRACYIFAKFXRNDDYFMQGSTY
 \\
 2   & CSEZMVWZLRMTDGBCPFASTRZMZVYMDVMADPKPAAUAXDVKMFVJEZEWSCNTXJICYSPDLXBTHEWPBJVWVXKDCSIFAMFXBOHMEZTQGTJR \\
 3   & OOERJVDAKRCWDAUIPFESTUILVBLMHEMAFPKZRAIFJQVVMVFZLSTTNUATMJICYUSXLRBUPVQPWBFWAXRECKIFAMFORODFYFKODSTY \\
 4   & IYZQQUTVKRIWEYLCECQUTUZYVYLZFLMQMOKTAYVMHJILMXVJDVRCUTXJSHWWYWVXQSRUHIKUXAKSPYGACEVDALUOEGRFIXKWFWMK \\
 5   & OYEQFRDALRTWEXBIPIZQAWSLKXIMRHMMUPGVAAWFXDKVMPNEQZTGUUXJMKMCMQSXWFBPHLQJJLFGAXRACYIFAIFXRRDDYFKUGTKY \\
 6   & CHNQRHAZYNMTDRWCPFASGUSYZVLMDJIQUPKIOCXAHDVNHFVAQZEBSULTCNICYUPNYVBTHEWPJJVAVXRRLIHFAIFXTOVDEZGUGYKR \\
 7   & JCEZYRWLPRENDCWOQIZGFUSXBYIURHMSUHGZGMQAWQOVZPSTCZRRQTYZXPPSAUPNWFDUHLLJVLFVAQLTCBIVEIFXBRLWARGORRKH \\
 8   & JKFRYVTFFRDEDCHIBIVJGBFFIVZMXKMAMOYHWAOAHDTWAFDJQDGGXTXZCRZCAWYXNXBWHHWCJAVKEURALANDENQOWKTJIZSSRRUA \\
 9   & HYAFWFHZPRKWEYBQEFEQRQIYLYLZXLMQUPYTSZVAHPKLDXFJQVRCXLMJSHIWYXVJQEBUMIKWQAKSIQBACIRDALAYEGLWIDKWVWOF \\
10   & NXQHYVUOPROGEDFOQQOQGUGPKWRMDYUHUGXXAAVAHQTLMSVADZYYXPEJSHISMTFPWXBGCHRZIYXBVRLTCYAFANFMRRXGQFEGDMDN \\ \hline
\end{tabular}
}
}
\end{center}
\end{table}

\begin{figure}[htp]
\begin{center}
\resizebox{0.9\textwidth}{!}{\includegraphics{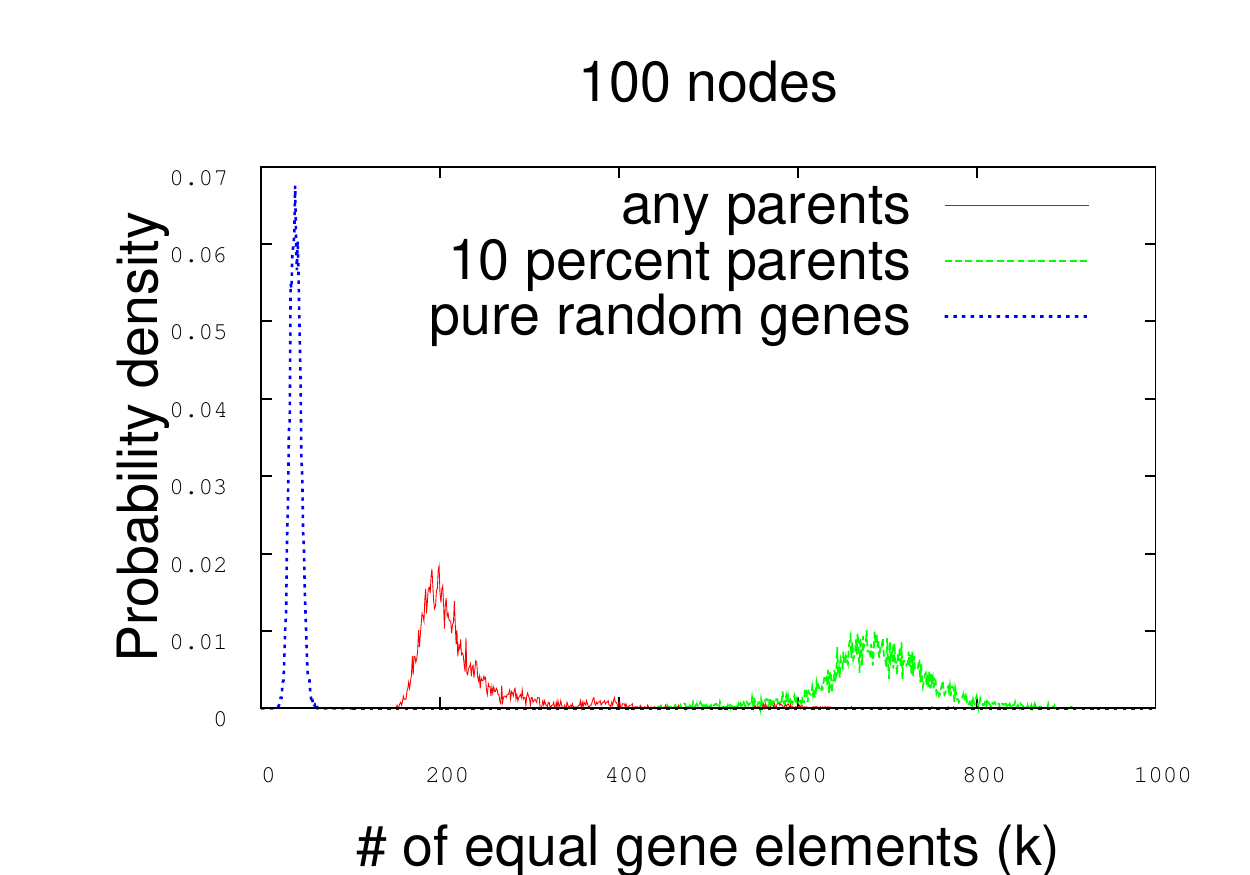}} \\ 
\resizebox{0.9\textwidth}{!}{\includegraphics{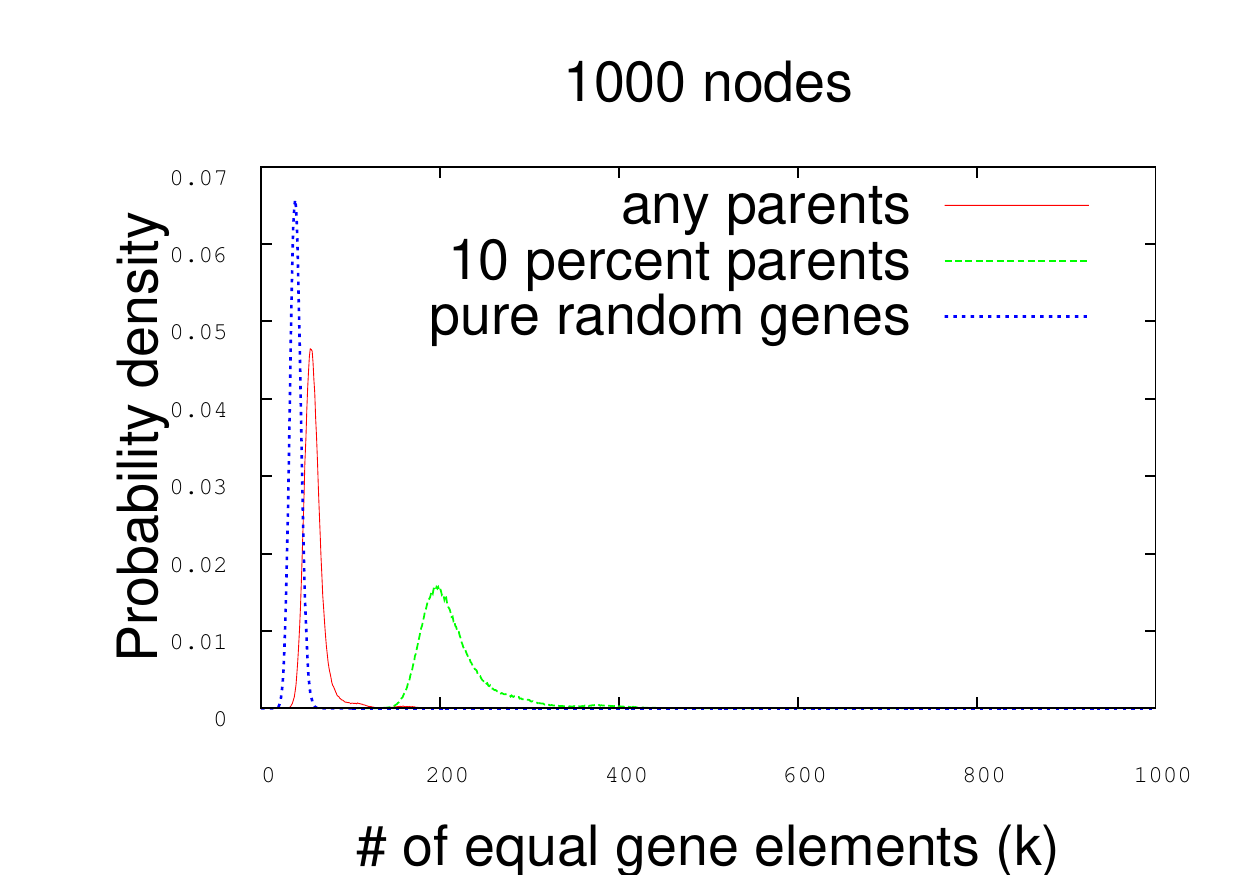}} 
\end{center} 
\caption{
Distribution of average pairwise number of equal gene elements for simulations of 100 and 1000 nodes.
The number of equal elements has a binomial distribution.
The left (blue) graph is for completely random genes. 
Note that all the nodes within a population of 100 nodes have significantly correlated genes.
The simulation with 1000 nodes gives similar correlation only for about 50 percent of the population of nodes.
However, if only 10 percent of the nodes can be parents, the nodes are always significantly correlated.
}
\label{fig:corel1}
\end{figure}

Figure \ref{fig:corel1} shows that for 100 nodes and only 10 possible parents, the genetic diversity is significant larger than 50 percent (about 700 out of 1000 corresponding gene elements are equal). 
This is due to ''inter-breeding'' between ancestors and descendants.
The tendency of an upper tail for the other similar distributions in Figure \ref{fig:corel1} has the same explanation. 

\begin{figure}[htp]
\begin{center}
\resizebox{0.99\textwidth}{!}{\includegraphics{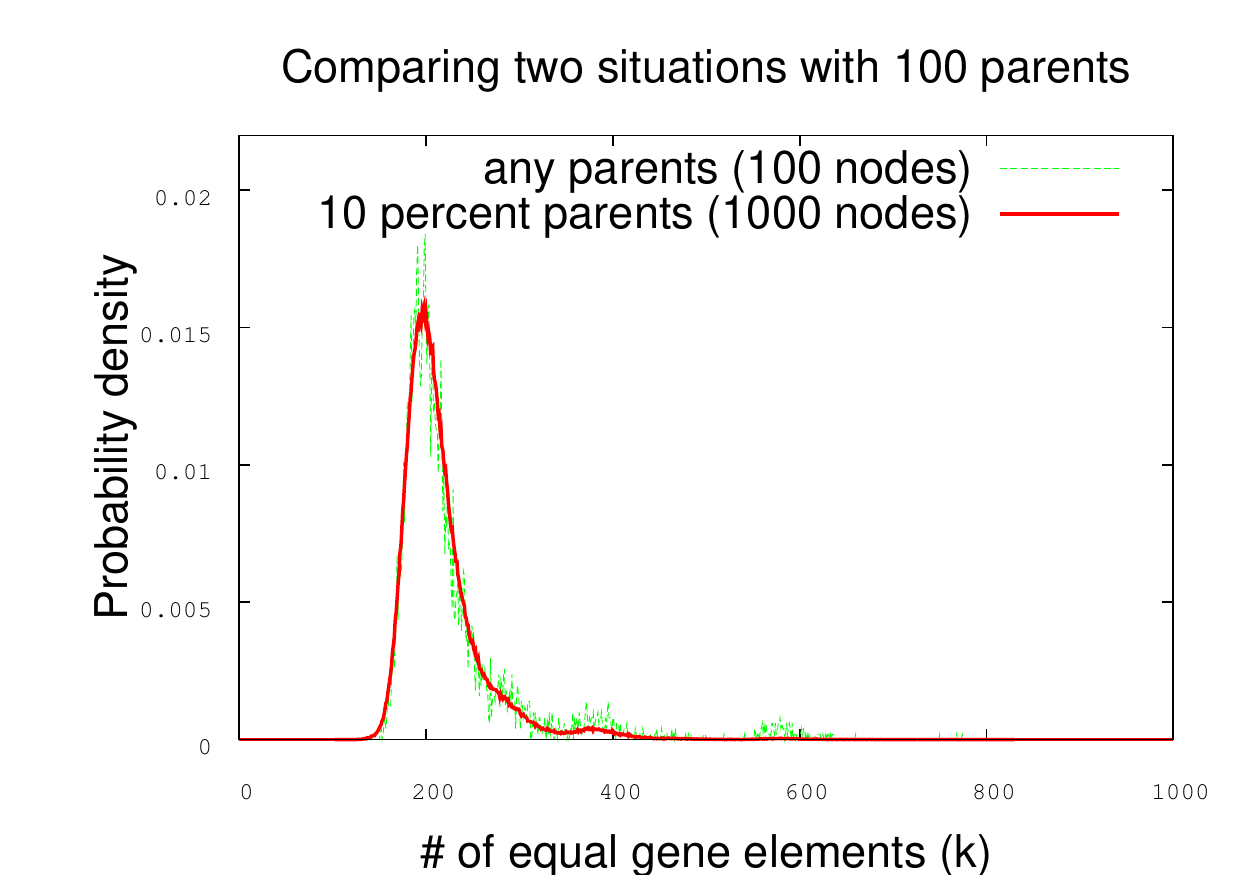}} \\
\end{center}
\caption{Comparison of two populations with same number
of potential parents (100 and 1000 nodes).} 
\label{fig:corel_1_2}
\end{figure}

Figure \ref{fig:corel_1_2} shows a comparison between the distribution of equal gene elements in two simulations of populations of respectively 100 and 1000 nodes.
Any node can be parent in the 100 nodes population whereas only 10 percent of the nodes can be parent in the 1000 node population. 
Hence the number of parents are equal for these two populations(100 potential parents).
Figure \ref{fig:corel_1_2} illustrates that the set of parents defines the gene statistics.
However, the extra offspring contribute to make many small variations (smooth distribution).

\begin{figure}[htp]
\begin{center}
\resizebox{0.99\textwidth}{!}{\includegraphics{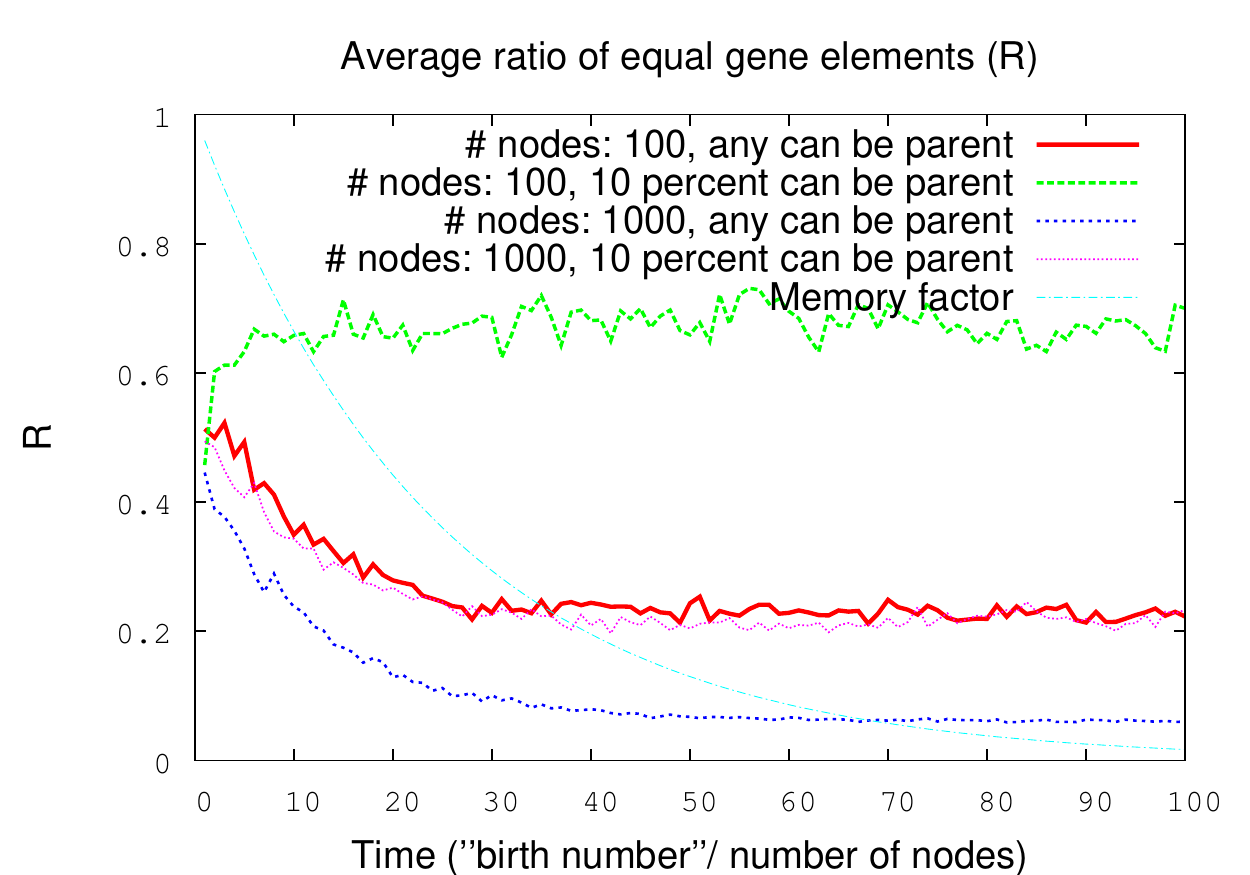}} 
\end{center}
\caption{Time evolution of average number of equal gene elements.
The birth rate per node is constant  
(i.e.\ the time for a birth in a 100 node network is 1/10 of the time in a 1000 node network).
A birth process generates a new gene sequence including 0.4 percent mutations. 
The ''memory factor'' $(1 - 0.04)^{t} = \exp(\mu t)$, where $\mu = \log 0.96$,  indicates the probability for a gene element to survive up to time $t$. 
}
\label{fig:correlation_file}
\end{figure}

Figure \ref{fig:correlation_file} shows the time development of the average number of equal gene elements for the above simulations of 100 and 1000 nodes. 
It shows, as expected, that this time development is equivalent for node populations with the same number of possible parents.  
Figure \ref{fig:corel_1_2} shows the distribution of the number of equal nodes at the end of the simulation.
Figure \ref{fig:correlation_file} also shows a ''memory factor''  $(1 - 0.04)^{t}$, where $t$ is time.
It indicates the probability for a gene element to survive (i.e.\ to avoid mutation) after a number of births given by the time parameter $t$.

\subsection{Transfer of secret key}
\label{se:transfer_of_secret_key}
The following elaboration shows the potential for using correlation between genes to code messages between nodes.
Assume a node X transmits a gene (sub-)sequence to another node Y within a network as above.
For each gene element $x = G^{X}_{i}$ which X transmits, there corresponds a statistically (positively) correlated element $y = G^{Y}_{i}$ in Y.
Figure \ref{fig:corel1} indicates this correlation.
Each gene element has equal probability in this case.
Assume X applies a ''code table'' mapping gene elements $g: \nucleus \rightarrow \nucleus$ before transmission.
The receiving node Y can estimate this mapping which can serve as an encryption key.
The mapping $g$ may be restricted to be a one-to-one (disambiguation) mapping given by a permutation of the set $\nucleus = \{ A,B,C,\ldots,Z \}$.
Note that there are here $26 ! = 1 \cdot 2 \cdot 3 \cdot \ldots \cdot 25 \cdot 26$ such possible permutations.

Table \ref{tab:codes} shows an example of the conditional distribution of the received code given the original code. 
\begin{table}[htp]
\caption{
Conditional distribution (percent) of gene element in receiving node Y given corresponding gene element in the node X.
Each row here represents such a distribution.
}
\label{tab:codes}
\begin{center}
\vspace{2mm}
\resizebox{\textwidth}{!}{
\begin{tabular}{r|rrrrrrrrrrrrrrrrrrrrrrrrrr|}
 & A& B& C& D& E& F& G& H& I& J& K& L& M& N& O& P& Q& R& S& T& U& V& W& X& Y& Z\\ \hline
 & \multicolumn{26}{|l|}{\mbox{}}    \\ 
A& 0& 0& 2&54& 2& 0& 0& 5& 0& 2& 2& 2& 7& 0& 0& 2& 2& 7& 0& 2& 0& 2& 2& 2& 0& 0\\    
B& 6& 3& 3& 0&44& 3& 0& 3& 3& 3& 3&11& 6& 0& 0& 0& 0& 0& 0& 0& 0& 0& 0& 3& 3& 8\\    
C& 6& 3& 3& 6& 0&54& 0& 3& 3& 3& 0& 0& 0& 3& 0& 0& 0& 6& 0& 0& 0& 9& 0& 3& 0& 0\\    
D& 0& 3& 0& 5& 3& 5&45& 0& 0& 3& 0& 3& 3& 0& 0& 0&10& 3& 3& 5& 3& 3& 0& 0& 5& 3\\    
E& 0& 0& 4& 4& 0& 9& 0&49& 2& 4& 0& 2& 6& 0& 0& 4& 0& 0& 0& 0& 9& 2& 0& 0& 0& 4\\    
F& 3& 3& 0& 3& 5& 0& 5& 3&62& 5& 0& 0& 0& 0& 3& 5& 0& 0& 3& 0& 0& 3& 0& 0& 0& 0\\    
G& 3& 3& 3& 0& 0& 0& 0& 6& 0&58& 6& 0& 0& 3& 3& 0& 0& 0& 0& 0& 6& 0& 3& 6& 0& 3\\    
H& 0& 0& 0& 3& 9& 0& 0& 6& 3& 3&40& 0& 6& 3& 9& 3& 3& 0& 3& 0& 3& 3& 3& 3& 0& 0\\   
I& 6& 2& 4& 0& 4& 0& 2& 2& 4& 0& 0&46& 2& 2& 2& 2& 4& 2& 2& 2& 6& 4& 0& 2& 2& 0\\   
J& 3& 0& 3& 3& 0& 0& 0& 0& 7& 7& 0& 0&48& 3& 0& 3& 3& 0& 3& 3& 0& 0& 0& 0& 7& 3\\   
K& 0& 0& 0& 6& 0& 0& 6& 0& 0& 3& 0& 3& 3&59& 3& 3& 0& 3& 3& 3& 0& 3& 0& 0& 0& 3\\   
L& 0& 5& 5& 0& 2& 0& 2& 2& 2& 0& 5& 0& 0& 2&52& 0& 2& 5& 0& 2& 0& 2& 2& 5& 0& 2\\   
M& 3& 0& 3& 3& 0& 0& 5& 3& 0& 3& 3& 3& 3& 5& 3&51& 0& 3& 5& 3& 3& 3& 0& 0& 0& 0\\   
N& 3& 0& 0& 0& 3& 0& 3& 0& 0& 3& 0& 3& 0& 0& 3& 3&52& 3& 0& 3& 0& 3& 3& 0& 0&13\\   
O& 0& 0& 0& 5& 3& 3& 3& 5& 3& 5& 0& 0& 5& 0& 0& 3& 5&48& 0& 3& 5& 3& 3& 0& 3& 0\\   
P& 0&10& 0& 2& 5& 0& 0& 0& 2& 2& 0& 0& 2& 2& 0& 2& 2& 0&57& 5& 0& 5& 0& 0& 0& 2\\   
Q& 2& 2& 7& 7& 2& 0& 2& 0& 0& 0& 5& 7& 0& 0& 2& 5& 0& 5& 2&42& 5& 0& 5& 0& 0& 0\\   
R& 2& 0& 5& 5& 0& 2& 2& 5& 5& 2& 0& 0& 2& 0& 0& 5& 2& 5& 2& 0&44& 2& 0& 0& 0& 7\\   
S& 6& 3& 0& 6& 3& 3& 3& 3& 0& 3& 3& 0& 3& 0& 3& 6& 6& 0& 0& 3& 3&40& 0& 0& 3& 3\\   
T& 0& 3& 3& 3& 0& 0& 6& 0& 0& 3& 0& 3& 0& 3& 3& 0& 0& 0& 0& 3& 9& 3&53& 0& 0& 6\\   
U& 0& 0& 0& 0& 0& 0& 0& 3& 5& 0& 0& 0& 5& 0& 3& 0& 3& 0& 3& 3& 3& 0& 0&59& 5&10\\   
V& 3& 0& 0& 6& 3& 0& 0& 6& 0& 0& 0& 3& 3& 3& 0& 0& 0& 3& 3& 0& 3& 3& 0& 0&62& 0\\   
W& 4& 0& 2& 0& 0& 2& 6& 0& 2& 4& 0& 4& 2& 0& 0& 0& 0& 2& 0& 2& 4& 2& 0& 4& 2&56\\   
X&53& 0& 6& 0& 0& 0& 3& 0& 0& 0& 0& 0& 0& 6& 0& 0& 0&12& 0& 3& 3& 3& 0& 9& 3& 0\\   
Y& 0&46& 0& 3& 0& 8& 3& 3& 5& 0& 5& 0& 3& 0& 3& 8& 0& 3& 3& 3& 0& 3& 0& 3& 0& 3\\   
Z& 0& 0&57& 0& 0& 0& 0& 0& 3& 6& 0& 0& 3& 3& 3& 6& 0& 3& 0& 0& 9& 9& 0& 0& 0& 0\\ \hline 
\end{tabular}
}
\end{center}
\end{table}
The first line of Table \ref{tab:codes} indicates that
the code table (function) maps A to D 
($g : \mbox{A} \rightarrow \mbox{D})$.
The second line similarly indicates $g \mbox{B} \rightarrow \mbox{E})$.
The three bottom lines indicate $g : \mbox{X} \rightarrow \mbox{A}$, $g : \mbox{Y} \rightarrow \mbox{B}$ and  $g : \mbox{Z} \rightarrow \mbox{C}$.

Let $X$ and $Y$ denote any two nodes.
Note that for $i \ne j$ any two different gene elements, $G_{i}^{X}$ and $G_{j}^{Y}$ are statistically uncorrelated. 
Assume that the mapping  \mbox{$\Phi : \{1,2,\ldots, n \} \rightarrow \{1,2,\ldots, n \}$} defines a permutation of the set of integers $1,2,\ldots,n$.  
I.e.\ $( \Phi(1),\Phi(2),\ldots,\Phi(n) )$ is a permutation of $(1,2,\ldots,n)$. 
Assume the node $X$ transmits the gene elements $x _{i} = G_{i}^{X}$, $i = 1,2,\ldots,n$ 
to the node $Y$ after translating then through the code table (mapping) $g$
as described above.
Assume $X$ also permutes the order of elements before transmitting it to node $Y$ translated by a code table (or function) $g$.
I.e.\ $Y$ receives the gene codes $y_{i} = g(G_{\phi(i)}^{X})$,
 $i = 1,2,\ldots,n$.
$Y$ holds the potential to recognise both the permutation $\phi$ and the code table $g$ due to its genetic relation  (code sequence correlation) with $X$.
The following outline shows this potential. Consider Table \ref{tab:codes}.
Each row represents an estimate $p(y \mid x)$ of the conditional probability distribution of a translated gene element $y = G_{i}^{Y} \in \nucleus$ in the node $Y$ given the corresponding element $x = G_{i}^{X} \in \nucleus$ in the node $X$(note: $\nucleus = \{ A,B,C, \ldots, Z \}$). 
Let $p(y \mid x)$ represent the element given by the row $x \in \nucleus$ and column $y \in \nucleus$ of Table \ref{tab:codes}. 
The permutation $\phi$ tend to minimise the entropy measure $H_{\phi}$:
\begin{equation}
\label{eq:entropy1}
H(Y \mid X_{\phi}) = 
          - \sum_{i=1}^{n} p(y_{i} \mid x_{\phi(i)}) \log p( y_{i} \mid x_{\phi(i)})
\end{equation}
It similarly maximises the mutual information:
\begin{equation}
\label{eq:mutual_information1}
I(X_{\phi};Y) = \sum_{x_{\phi(i)},y_{i}} p(x_{\phi(i)}, y_{i}) \log \frac{p(x_{\phi(i)}, y_{i})} {p(x_{\phi(i)}) p(y_{i})}
\end{equation}

Hence, $Y$ may sort out likely candidates for the original permutation $\phi$ decided by the node $X$. 
Application of Equations \ref{eq:entropy1} and \ref{eq:mutual_information1} can significantly save computational cost.
The unconditional distribution of the value of a gene element $G_{i}^{X}$ is for example constant $p(x) = 1/M$ where $M$ is the number of possible values of a gene element ($M = 26$ for the present example). 
Table \ref{tab:codes} demonstrates a possible computational simplification.
The actual permutation $\phi$ is the one that gives (only) one value significant larger than the others along each row.
The resulting ''minimum entropy table'' then gives the code table $g$.

Restriction of possible permutations and code tables also reduces potential computational cost.
However, it increases the leakage of information to the environment.
Hash values from the node $X$ may help $Y$ finally to find the actual permutation among a limited number of likely candidates.

\subsection{Recognition as password cracking}
\label{se:password_cracking}
Assume two nodes X and Y share some ancestors (are relatives) and that corresponding gene elements are equal with probability $p = 1/3$ as compared to a not relative where the probability of equal gene element would be $p = 1/26$ (note that a gene element can have 26 possible values). 
Assume X asks Y to guess the value of a number of its (X's) gene elements (for example the $n$ first gene elements of X).
One may look at this situation as if Y has to guess a password (or a set of possible passwords).
X may recognise Y as ''relative'' if it is clever to guess requested passwords (either measured by response time or by ratio of success assuming X gives Y several fake alternative hash values for the passwords so a non-relative will more often guess wrong password as compared to a relative).

One may intuitively believe that a probability of $p = 1/3$ to guess correctly each character in a password, does not help much to guess a password of many characters.
The examples below may help perception.
Assume, for simplicity, that passwords are only two characters long.
This gives a search space consisting of $26 \times 26 = 676$ possible combinations of characters in the range A,B,\ldots,Z (26 possible letters).
Assume the two first gene elements of Y are 'AA' (i.e.\ $G_{1}^{Y} = G_{2}^{Y} = \mbox{A}$).
This means that   $P(G_{1}^{X} = \mbox{A} \mid G_{1}^{Y} = \mbox{A}) = 1/3$ and $P(G_{1}^{X} = \mbox{S} \mid G_{1}^{Y} = \mbox{A}) = 2/3 \times 1/25 = 2/75$ for any value of $S$ different from A.

\newcommand{\pp}{\ensuremath{\mathbf{pp}}}
\newcommand{\pq}{\ensuremath{\mathbf{pq}}}
\newcommand{\qp}{\ensuremath{\mathbf{qp}}}
\newcommand{\qq}{\ensuremath{\mathit{qq}}}

Table \ref{tab:password_probabilities1} shows that Y may utilise information in its genes to significantly restrict the search space for the correct password to find it with probability larger than 50 percent.
\begin{table}[htp]
\caption{Probabilities for the first two gene elements of X given that the corresponding gene elements of Y is 'AA'
(i.e.\ $G_{1}^{Y} = G_{2}^{Y} = \mbox{A}$).
$p = 1/3$ is the (conditional) probability for the first and second gene element of
X to be A (independently).
$q = 2/75$ is the probability any other value of these gene elements.
Left column and upper row respectively annotate values of first and second gene elements.
Combinations of $p$ and $q$ are products.
$pp = (1/3)^{2}$, $pq = 1/3 \times 2/75$ and $\qq = 2/75 \times 2/75$.
The sum of the products given by the union of the row and the column A ({\bf bold} letters) is 0.56.
This part of the search space thus has probability more than 0.5.
Note that the table can be looked at as a result from the matrix operation $[ p, q, q, \ldots, q ]^{t} [ p, q, q, \ldots, q ]$ between two vectors giving the (conditional) probability for each character A,B,...,Z. 
}
\label{tab:password_probabilities1}
\vspace{1mm}
\begin{center}
\resizebox{\textwidth}{!}{
\begin{tabular}{r|c|ccccccccccccccccccccccccc|}
 & A& B& C& D& E& F& G& H& I& J& K& L& M& N& O& P& Q& R& S& T& U& V& W& X& Y& Z\\ \hline 
 & \mbox{} & \multicolumn{25}{l|}{\mbox{}}    \\ 
A&\pp&\pq&\pq&\pq&\pq&\pq&\pq&\pq&\pq&\pq&\pq&\pq&\pq&\pq&\pq&\pq&\pq&\pq&\pq&\pq&\pq&\pq&\pq&\pq&\pq&\pq\\ \hline
B&\qp&\qq&\qq&\qq&\qq&\qq&\qq&\qq&\qq&\qq&\qq&\qq&\qq&\qq&\qq&\qq&\qq&\qq&\qq&\qq&\qq&\qq&\qq&\qq&\qq&\qq\\    
C&\qp&\qq&\qq&\qq&\qq&\qq&\qq&\qq&\qq&\qq&\qq&\qq&\qq&\qq&\qq&\qq&\qq&\qq&\qq&\qq&\qq&\qq&\qq&\qq&\qq&\qq\\    
D&\qp&\qq&\qq&\qq&\qq&\qq&\qq&\qq&\qq&\qq&\qq&\qq&\qq&\qq&\qq&\qq&\qq&\qq&\qq&\qq&\qq&\qq&\qq&\qq&\qq&\qq\\    
E&\qp&\qq&\qq&\qq&\qq&\qq&\qq&\qq&\qq&\qq&\qq&\qq&\qq&\qq&\qq&\qq&\qq&\qq&\qq&\qq&\qq&\qq&\qq&\qq&\qq&\qq\\    
F&\qp&\qq&\qq&\qq&\qq&\qq&\qq&\qq&\qq&\qq&\qq&\qq&\qq&\qq&\qq&\qq&\qq&\qq&\qq&\qq&\qq&\qq&\qq&\qq&\qq&\qq\\    
G&\qp&\qq&\qq&\qq&\qq&\qq&\qq&\qq&\qq&\qq&\qq&\qq&\qq&\qq&\qq&\qq&\qq&\qq&\qq&\qq&\qq&\qq&\qq&\qq&\qq&\qq\\    
H&\qp&\qq&\qq&\qq&\qq&\qq&\qq&\qq&\qq&\qq&\qq&\qq&\qq&\qq&\qq&\qq&\qq&\qq&\qq&\qq&\qq&\qq&\qq&\qq&\qq&\qq\\    
I&\qp&\qq&\qq&\qq&\qq&\qq&\qq&\qq&\qq&\qq&\qq&\qq&\qq&\qq&\qq&\qq&\qq&\qq&\qq&\qq&\qq&\qq&\qq&\qq&\qq&\qq\\    
J&\qp&\qq&\qq&\qq&\qq&\qq&\qq&\qq&\qq&\qq&\qq&\qq&\qq&\qq&\qq&\qq&\qq&\qq&\qq&\qq&\qq&\qq&\qq&\qq&\qq&\qq\\    
K&\qp&\qq&\qq&\qq&\qq&\qq&\qq&\qq&\qq&\qq&\qq&\qq&\qq&\qq&\qq&\qq&\qq&\qq&\qq&\qq&\qq&\qq&\qq&\qq&\qq&\qq\\    
L&\qp&\qq&\qq&\qq&\qq&\qq&\qq&\qq&\qq&\qq&\qq&\qq&\qq&\qq&\qq&\qq&\qq&\qq&\qq&\qq&\qq&\qq&\qq&\qq&\qq&\qq\\    
M&\qp&\qq&\qq&\qq&\qq&\qq&\qq&\qq&\qq&\qq&\qq&\qq&\qq&\qq&\qq&\qq&\qq&\qq&\qq&\qq&\qq&\qq&\qq&\qq&\qq&\qq\\    
N&\qp&\qq&\qq&\qq&\qq&\qq&\qq&\qq&\qq&\qq&\qq&\qq&\qq&\qq&\qq&\qq&\qq&\qq&\qq&\qq&\qq&\qq&\qq&\qq&\qq&\qq\\    
O&\qp&\qq&\qq&\qq&\qq&\qq&\qq&\qq&\qq&\qq&\qq&\qq&\qq&\qq&\qq&\qq&\qq&\qq&\qq&\qq&\qq&\qq&\qq&\qq&\qq&\qq\\    
P&\qp&\qq&\qq&\qq&\qq&\qq&\qq&\qq&\qq&\qq&\qq&\qq&\qq&\qq&\qq&\qq&\qq&\qq&\qq&\qq&\qq&\qq&\qq&\qq&\qq&\qq\\    
Q&\qp&\qq&\qq&\qq&\qq&\qq&\qq&\qq&\qq&\qq&\qq&\qq&\qq&\qq&\qq&\qq&\qq&\qq&\qq&\qq&\qq&\qq&\qq&\qq&\qq&\qq\\    
R&\qp&\qq&\qq&\qq&\qq&\qq&\qq&\qq&\qq&\qq&\qq&\qq&\qq&\qq&\qq&\qq&\qq&\qq&\qq&\qq&\qq&\qq&\qq&\qq&\qq&\qq\\    
S&\qp&\qq&\qq&\qq&\qq&\qq&\qq&\qq&\qq&\qq&\qq&\qq&\qq&\qq&\qq&\qq&\qq&\qq&\qq&\qq&\qq&\qq&\qq&\qq&\qq&\qq\\    
T&\qp&\qq&\qq&\qq&\qq&\qq&\qq&\qq&\qq&\qq&\qq&\qq&\qq&\qq&\qq&\qq&\qq&\qq&\qq&\qq&\qq&\qq&\qq&\qq&\qq&\qq\\    
U&\qp&\qq&\qq&\qq&\qq&\qq&\qq&\qq&\qq&\qq&\qq&\qq&\qq&\qq&\qq&\qq&\qq&\qq&\qq&\qq&\qq&\qq&\qq&\qq&\qq&\qq\\    
V&\qp&\qq&\qq&\qq&\qq&\qq&\qq&\qq&\qq&\qq&\qq&\qq&\qq&\qq&\qq&\qq&\qq&\qq&\qq&\qq&\qq&\qq&\qq&\qq&\qq&\qq\\    
W&\qp&\qq&\qq&\qq&\qq&\qq&\qq&\qq&\qq&\qq&\qq&\qq&\qq&\qq&\qq&\qq&\qq&\qq&\qq&\qq&\qq&\qq&\qq&\qq&\qq&\qq\\    
X&\qp&\qq&\qq&\qq&\qq&\qq&\qq&\qq&\qq&\qq&\qq&\qq&\qq&\qq&\qq&\qq&\qq&\qq&\qq&\qq&\qq&\qq&\qq&\qq&\qq&\qq\\    
Y&\qp&\qq&\qq&\qq&\qq&\qq&\qq&\qq&\qq&\qq&\qq&\qq&\qq&\qq&\qq&\qq&\qq&\qq&\qq&\qq&\qq&\qq&\qq&\qq&\qq&\qq\\    
Z&\qp&\qq&\qq&\qq&\qq&\qq&\qq&\qq&\qq&\qq&\qq&\qq&\qq&\qq&\qq&\qq&\qq&\qq&\qq&\qq&\qq&\qq&\qq&\qq&\qq&\qq\\ \hline    
\end{tabular}
}
\end{center}
\end{table}
The whole search space consists of 676 elements (possible combinations).
Hence an outsider (non relative) has to search through 338 combinations to find the combinations with 50 percent probability. 
However, Y may restrict its search to the 51 combinations with highest probability (row and column A).
Denote such a search space as a ''50 percent probability search space''.
A significant intuition here is that the cardinality of the 50 percent conditional probability search space scales with dimensions below the similar space for unconditional probabilities (or less restricted conditions). 

An identification procedure may include several such tests and an outsider may loose the game of guessing passwords by processing too slow or via giving wrong answer too frequent (provided he is given several alternative hash values for the search).  
 
The following example generalises the one above.
Assume now that X asks Y to guess the value of 10 of its (X's) gene elements (for example the 10 first gene elements of X).
The node X may deliver to Y a hash value of these gene elements so Y can check if it guesses it correct (or alternatively give a set of possible hash values so Y may risk to give wrong answer).
The sequence of these 10 gene elements forms a key (or ''password'') which Y is asked to resolve (''crack''). 
The task of Y is similar to crack a realistic password given a (Unix) password file.
Node Y can reply to node X with another hash value.

\begin{figure}[htp]
\begin{center}
\resizebox{0.8\textwidth}{!}{\includegraphics{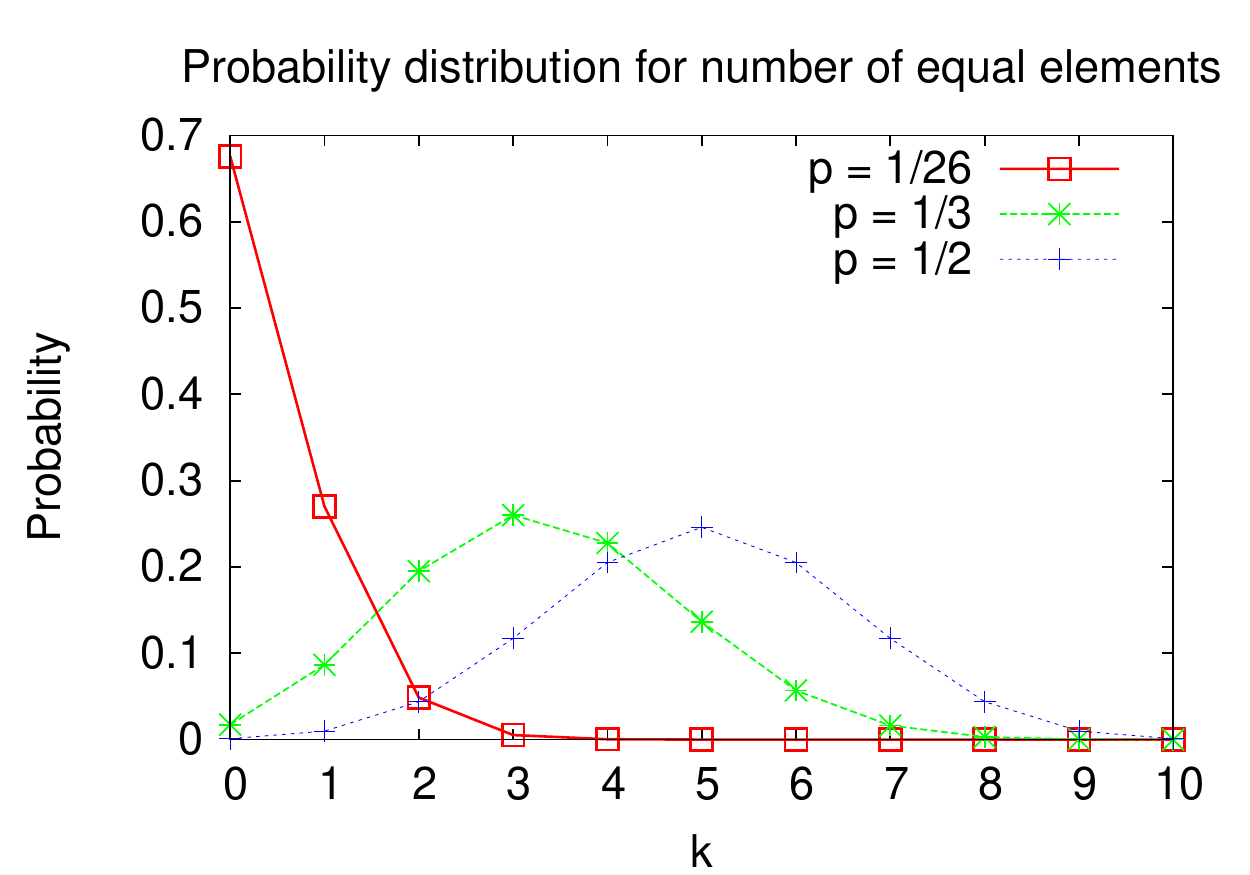}}
 
\end{center}
\caption{Probability of equal corresponding gene elements.}
\label{fig:binomial1}
\end{figure}

The knowledge that $p = 1/3$ gives the node Y also here the opportunity to define a probability measure on the set of all possible keys (or ''passwords'').
Let $k$ denote the number of equal gene elements in the corresponding sequence of gene elements of the nodes X and Y ($0 \le k \le 10$).
$k$ has the following (binomial) probability mass distribution:
\begin{equation}
f(k;n,p) = {n \choose k} p^{k} (1 - p)^{n - k} 
\end{equation}
where $n = 10$ and $p$ is the probability of pair-wise equal gene element.
Figure \ref{fig:binomial1} shows the distribution of number of
Pair-wise equal elements for $p = 1/26, 1/3, 1/2$.

Note that in the case of $p = 1/26$ (not relatives)
the probability $P(C)$ for any random combination $C$ of gene elements is
\begin{equation} 
    P(C) =
    \frac{ {n \choose k} p^{k} (1 - p)^{n-k} } 
         { {n \choose k} 25^{n - k} } 
  = \frac{1}{26^{n}}
\end{equation} 
This means (as expected) that all combinations have the same probability for being the correct key (''password'') for not related nodes.
Hence, for $p = 1/26$, in order to find the correct key with more than 50 percent probability, Y must test $26^{10}/2$ combinations of gene element values.

Figure \ref{fig:binomial1} shows that for $p = 1/3$ (relative), more than half of the probability mass is for $k = 3, 4, 5$.
Equation \ref{eq:brute_force1} gives the general formula for the total number of combinations $N_{k}$ for a given number $k$ of matching elements.
\begin{equation}
\label{eq:brute_force1}
N_{k} = {n \choose k} (I - 1)^{n-k}
\end{equation}
$I$ is number of possible values for each gene element (in this case $I = 26$).
This gives that the total number of tests (for $k = 3, 4, 5$ and $p = 1/26$) is 
${n \choose 3} 25^{7} +  {n \choose 4} 25^{6} +  {n \choose 5} 25^{5}$.
Hence a non-relative has to process (test) in average about 1000 times more different keys (combinations) as compared to a relative to find the correct key with a probability larger than 50 percent.
This difference (ratio) increases with increased value of $p$ (for relative), increased number of possible gene elements and increased  key length.
The ratio is for example more than a million for $I = 1024$(and $n = 10$ and $p = 1/3$ as above).

\section{Discussion}
\label{se:contribution}

The present approach can be redundant and complementary to centrally organised trusted components such as for Public Key Infrastructure \cite{He_1998_1012}.
Centrally organised and designed security systems typically lack robustness, distributability and autonomy.
They require correct implementation and management \cite[]{Somayaji1997}.
Hence one seeks alternatives for systems to operate in hostile or uncontrolled environments of for example users not caring for security.

Section \ref{se:example} illustrates by examples the potential to improve security in sensor networks where the nodes possess interrelated individual ''genetic'' codes with a restricted lifetime.
These examples are only meant to communicate ideas and to show potentials for control/regulation of statistical correlation between gene codes.
The nodes in these examples carry a dynamic (time changing) ''gene pool'' which can be looked at as a distributed common secret for recognition and protection of communication. 
This method of ''genetic protection'' has similarities to application of threshold cryptography where nodes share a cryptographic secret without each storing the entire information of it \cite{Shamir1979}.
The present simulation examples show protection without direct adaptation to specific adversaries (''patogens'').
The protection method is in this way similar to innate (static) immune systems.
However, assume an adversary manage to obtain genetic information from the gene pool and to use it for intrusion.
If it fails to fit into the phylogenetic three of the gene pool, or it appears as a clone of a member of the network, then the network may regenerate the part of the gene pool which makes the intruder able to enter the network.
This gives adaptation to specific intruders (cf the concept of ''adaptive immune systems'').

Genetic protection also has similarities to application of chaos cryptography \cite{Kinzel2002,Kanter2008a} where nodes obtain common secrets via a synchronisation process and which they can use to protect communication.
A node must participate in the synchronisation to obtain the common secret (encryption key) which is time varying.
A significant difference here (from the present approach) is that application of chaos synchronisation requires frequent communication/updates. 

Note that both chaos cryptography and numerically based cryptography \cite{diffie76new} over public channels imply active participation in a communication for a node to obtain cryptographic material. 
The examples above are similar in this way.
A node which is passive long enough, will fall outside the communication.

The present ideas have the justification from potential advantages in given situations.
These situations can be defined by for example low communication bandwidth, periods with unidirectional communication and attacks on centralised security systems.
Risk for loss of data is often a special issue for sensor networks.
Nodes in sensor networks may store valuable data and redundant security systems may help to secure these before a possibly expendable sensor system halts.
Sensor systems typically collect data which are available in the environment.
Hence protection of these data from being available for outsiders may have little meaning.
Protection of functionality may therefore be a main focus for security within sensor networks.

\bibliographystyle{elsart-num}
% \bibliographystyle{elsart-harv}

%\bibliography{rkffi}

\begin{thebibliography}{10}
\expandafter\ifx\csname url\endcsname\relax
  \def\url#1{\texttt{#1}}\fi
\expandafter\ifx\csname urlprefix\endcsname\relax\def\urlprefix{URL }\fi

\bibitem{Akyildiz2002}
I.~F. Akyildiz, W.~Su, Y.~Sankarasubramaniam, E.~Cayirci, A survey on sensor
  networks, IEEE Communications Magazine (2002) 102--114.
\newline\urlprefix\url{http://citeseer.ist.psu.edu/akyildiz02survey.html}

\bibitem{Aickelin2007}
U.~Aickelin, J.~Greensmith, J.~Twycross, Immune system approaches to intrusion
  detection - a review (2004).
\newline\urlprefix\url{http://citeseerx.ist.psu.edu/viewdoc/summary?doi=10.1.1%
.1.8176}

\bibitem{Jungwon2007}
J.~Kim, P.~J. Bentley, U.~Aickelin, J.~Greensmith, G.~Tedesco, J.~Twycross,
  Immune system approaches to intrusion detection --- a review, Natural
  Computing: an international journal 6~(4) (2007) 413--466.

\bibitem{Gordon2006}
L.~A. Gordon, M.~P. Loeb, W.~Lucyshyn, R.~Richardson, Computer crime and
  security survey, Tech. rep., Computer Security Institute (2006).

\bibitem{Shafi2007}
K.~Shafi, H.~A. Abbass, Biologically-inspired complex adaptive systems
  approaches to network intrusion detection, Inf. Secur. Tech. Rep. 12~(4)
  (2007) 209--217.

\bibitem{Kitano2007}
H.~Kitano, Towards a theory of biological robustness 3:137, Molecular Systems
  Biology.

\bibitem{Kitano2004}
H.~Kitano, Biological robustness, Nature Reviews Genetics 5 (2004) 826--837.

\bibitem{Stelling2004}
J.~Stelling, U.~Sauer, Z.~Szallasi, I.~F. Doyle, J.~Doyle, Robustness of
  cellular functions, Cell 118 (2004) 675--685.

\bibitem{Tampesti2007}
G.~Tampesti, Biological inspiration in the design of computing systems, in:
  Proceedings of the IEEE, Vol.~95, 2007, pp. 463--464.

\bibitem{Somayaji2007}
A.~Somayaji, Immunology, diversity, and homeostasis: The past and future of
  biologically inspired computer defenses, Inf. Secur. Tech. Rep. 12~(4) (2007)
  228--234.

\bibitem{Cohen1987}
F.~Cohen, Computer viruses: theory and experiments, Comput. Secur. 6~(1) (1987)
  22--35.

\bibitem{Guinier1989}
D.~Guinier, Biological versus computer viruses, SIGSAC Rev. 7~(2) (1989) 1--15.

\bibitem{Li2007}
J.~Li, P.~Knickerbockera, Functional similarities between computer worms and
  biological pathogens, Computers \& Security 26~(4).

\bibitem{Ibrahim2005}
S.~Ibrahim, M.~A. Maarof, A review on biological inspired computation in
  cryptology, Jurnal Teknologi Maklumat 1~(17) (2005) 90--98.

\bibitem{He_1998_1012}
Q.~He, K.~P. Sycara, Z.~Su, {A Solution to Open Standard of PKI}, in: ACISP
  '98: Proceedings of the Third Australasian Conference on Information Security
  and Privacy, Springer-Verlag, London, UK, 1998, pp. 99--110.

\bibitem{Somayaji1997}
A.~Somayaji, S.~Hofmeyr, S.~Forrest, Principles of a computer immune system,
  in: NSPW '97: Proceedings of the 1997 workshop on New security paradigms,
  ACM, New York, NY, USA, 1997, pp. 75--82.

\bibitem{Shamir1979}
A.~Shamir, How to share a secret, Commun. ACM 22~(11) (1979) 612--613.
\newline\urlprefix\url{http://dx.doi.org/10.1145/359168.359176}

\bibitem{Kinzel2002}
W.~Kinzel, I.~Kanter, Neural cryptography, in: Proc. of the 9th International
  Conference on Neural Information Processing, Vol.~3, 2002, pp. 1351--1354.

\bibitem{Kanter2008a}
I.~Kanter, E.~Kopelowitz, W.~Kinzel, Public channel cryptography: Chaos
  synchronization and hilbert's tenth problem, Physical Review Letters 101~(8).
\newline\urlprefix\url{http://dx.doi.org/10.1103/PhysRevLett.101.084102}

\bibitem{diffie76new}
W.~Diffie, M.~E. Hellman, New directions in cryptography, IEEE Transactions on
  Information Theory IT-22~(6) (1976) 644--654.
\newline\urlprefix\url{citeseer.ist.psu.edu/diffie76new.html}

\bibitem{Crescenzo2007}
G.~D. Crescenzo, R.~Ge, G.~R. Arce, Threshold cryptography in mobile {\em ad
  hoc} networks under minimal topology and setup assumptions, Ad Hoc Networks
  5~(1) (2007) 63--75.

\bibitem{Yi2003moca}
S.~Yi, R.~Kravets, Moca: Mobile certificate authority for wireless ad hoc
  networks, in: {2nd Annual PKI Research Workshop Program (PKI 03}, 2003, pp.
  65--79.

\end{thebibliography}

\end{document}